\documentclass[twocolumn,secnumarabic,amssymb, nobibnotes, aps, pr,superscriptaddress,showpacs]{revtex4-1}

\usepackage[]{graphicx}
\usepackage[]{epstopdf}
\usepackage[]{amsmath}
\usepackage[]{mathtools}

\newcommand{\iu}{\mathrm{i}}
\newcommand{\Ham}{{\hat{\mathrm{H}}}}

\newcommand{\Ket}[1]{{\left| #1 \right>}}
\newcommand{\BraKet}[2]{\left< #1 | #2 \right>}

\begin{document}

\title{Anisotropic Excitons and their Contributions to Shift Current Transients in Bulk GaAs}%

\author{Reinold Podzimski}%
\affiliation{Department Physik and Center for Optoelectronics and Photonics Paderborn (CeOPP), Universit\"at Paderborn, Warburger Str. 100, D-33098 Paderborn, Germany}

\author{Huynh Thanh Duc}%
\affiliation{Ho Chi Minh City Institute of Physics, Vietnam Academy of Science and Technology, Mac Dinh Chi Street 1, District 1, Ho Chi Minh City, Vietnam}

\author{Torsten Meier}%
\affiliation{Department Physik and Center for Optoelectronics and Photonics Paderborn (CeOPP), Universit\"at Paderborn, Warburger Str. 100, D-33098 Paderborn, Germany}

\date{\today}%

\begin{abstract}
Shift current transient are obtained for near band gap excitation of bulk GaAs
by numerical solutions of the semiconductor Bloch equations in a basis obtained from a 14 band {\bf k}$\cdot${\bf p} model of the band structure.
This approach provides a transparent description of the optically induced excitations in terms of interband, intersubband, and intraband excitations
which enables a clear distinction between different contributions to the shift current transients
and fully includes resonant as well as off-resonant processes.
Using a geodesic grid in reciprocal space in our numerical solutions,
we are able to include the electron-hole Coulomb attraction in combination with our anisotropic three-dimensional band structure.
We obtain an excitonic absorption peak and an enhancement of the continuum absorption
and demonstrate that the excitonic wave function contains a significant amount of anisotropy.
Optical excitation at the excitonic resonance generates shift current transients of significant strength, however,
due to the electron-hole attraction the shift distance is smaller than for above band gap excitation.
We thus demonstrate that our approach is able to provide
important information on the ultrafast electron dynamics on the atomic scale.
\end{abstract}


\maketitle

\section{Introduction}

The optical excitation of non-centrosymmetric crystals can be used to generate photocurrents on ultrafast time scales without the need of an external bias.
As shown by Sipe and co-workers, the lack of inversion symmetry in zincblende III-V semiconductors results in a non-vanishing zero-frequency second-order optical susceptibility $\chi^{(2)}$
which corresponds to photocurrents that can be generated by optical excitation with a single frequency.\cite{Sipe_PRB_00, Nastos_PRB_06}
One can distinguish between three types of photocurrents:
(i) injection currents originating from non-symmetric electronic distributions in k-space after resonant above band gap excitation,
(ii) shift currents which are due to the spatial motion of optically excited carriers in real space after above band gap excitation,
and (iii) rectification currents that result from the non-resonant polarization generated for below band gap excitation.

Here, we investigate the ultrafast dynamics of bulk GaAs following the near-band gap excitation by femtosecond laser pulses and
focus on analyzing shift currents which are responsible for the bulk photo-voltaic effect.
Experimentally, shift currents have been investigated in bulk semiconductors\cite{Zhang_PRL_92,Dalba_PRL_95,Chen_PRL_97,Cote_JAP_02,Somma_PRL_14}
and semiconductor quantum wells\cite{Mark_JAP_05,Mark_JAP_06,Mark_PRB_07,Mark_JAP_09,Duc_PRB_2016}.
Previous theoretical research on shift currents was mainly performed in the frequency domain using a perturbative analysis of the light-mater interaction
to derive analytical expression for the considered nonlinear response in terms of matrix elements and resonance denominators.\cite{Nastos_PRB_06,Young_PRL_12,Young_PRL_13,Kral_JPCM_00,Brehm_JAC_14}
Using the semiconductor Bloch equations (SBE) in the basis {\bf k}$\cdot${\bf p} wave functions it is possible to obtain photocurrents directly
in the time-domain as was shown for the case of injection currents
\cite{Duc_PRL_10,Duc_PRB_10,Duc_PSS_11}
as well as for shift and rectification currents \cite{Podzimski_SPIE_15,Podzimski_SPIE_16,Duc_PRB_2016}.
This approach provides a transparent description of the optical excitations in terms of interband, intersubband, and intraband excitations,
allows to treat the light-matter interaction non-perturbatively, and provides good agreement with experimental results on
injection and shift currents of GaAs based quantum well systems, e.g., \cite{Duc_PRL_10,Duc_PRB_2016}.

Due to the tremendous numerical demands, excitonic effects have been neglected in most previous theoretical investigations of photocurrents in semiconductors.
Whereas this is justified for high above band gap excitations where excitonic effects have negligible contributions,
for near band gap excitations, the many-body Coulomb interaction, in particular the electron-hole attraction, strongly modifies
the optical response and therefore needs to be incorporated into the theoretical approach.
Often when many-body effects are considered simplified models for the band structure and electronic states,
e.g., isotropic and/or parabolic band structures and a small number of bands, are used
which significantly reduces the numerical requirements when solving the
SBE.\cite{Zimmermann_PRB_1997,Rudin_PRB_2002,Bhat_PRB_2005,Turkowski_PRB_2008,Turkowski_PRB_2009,Haug_Koch_2004}
However, since the $\chi^{(2)}$ photocurents in general originate from anisotropies of the band structure and/or of the optical matrix elements,
they cannot be described properly by models using isotropic band structure models.
Furthermore, shift currents, in particular, involve non-resonant excitations and can therefore not be adequately described in models that consider only bands
that are present at or near the band gap.\cite{Podzimski_SPIE_15,Duc_PRB_2016}

For the theoretical analysis of excitonic contributions to shift currents in bulk GaAs we therefore apply our 
approach of the SBE formulated in a basis of electronic eigenstates obtained from a 14 band {\bf k}$\cdot${\bf p} model.
The anisotropic electronic band structure and the electronic states are well suited to describe shift currents transients.\cite{Podzimski_SPIE_15,Podzimski_SPIE_16,Duc_PRB_2016}
For the incorporation of excitonic resonances we extend our previous analysis and include the Coulomb interaction in time-dependent Hartree-Fock approximation.
Due to the anisotropic band structure and matrix elements and the need to incorporate off-resonant excitations, which means that the
rotating-wave approximation cannot be applied, the solutions of the resulting SBE are numerically very demanding.
To obtain converged results we employ a geodesic grid in k-space which provides results of reasonable accuracy with a significantly smaller
numerical effort than a cartesian grid.

In Sec.~\ref{sec::Theory} we describe the fundamentals of our theoretical approach and the idea behind the use of the geodesic grid in our simulations.
The excitonic absorption spectrum and the anisotropic exciton wave function are presented in Sec.~\ref{sec::Exciton}.
Numerical results on shift current transients including excitonic effects are shown and discussed in Sec.~\ref{sec::Shift}.

\section{Theoretical Approach \& Numerical Challenges}
\label{sec::Theory}

The {\bf k}$\cdot${\bf p}-based extended Kane model is used for the calculation of the semiconductor band structure.
The extended Kane model is represented by a 14-band Hamiltonian,
\begin{equation}
\label{sec:Theory;eq::1}
\Ham_{14\times14} = 
\begin{pmatrix}
\Ham_{8c8c} & \Ham_{8c7c}  & \Ham_{8c6c}  & \Ham_{8c8v} & \Ham_{8c7v}  \\
\Ham_{7c8c} & \Ham_{7c7c}  & \Ham_{7c6c}  & \Ham_{7c8v} & \Ham_{7c7v}  \\
\Ham_{6c8c} & \Ham_{6c7c}  & \Ham_{6c6c}  & \Ham_{6c8v} & \Ham_{7c7v}  \\
\Ham_{8v8c} & \Ham_{8v7c}  & \Ham_{8v6c}  & \Ham_{8v8v} & \Ham_{8v7v}  \\
\Ham_{7v8c} & \Ham_{7v7c}  & \Ham_{7v6c}  & \Ham_{7v8v} & \Ham_{7v7v}  
\end{pmatrix}\text{,}
\end{equation}
which describes the band structure of zincblende crystals near the $\Gamma$-point, GaAs being one prominent example.\cite{Mayer_PRB_91,winkler_book_03,Trebin_PRB_79}
The Hamiltonian includes the split-off band $\Ket{7v}$, the highest valence band $\Ket{8v}$, the lowest conduction band $\Ket{6c}$, and higher conduction bands $\Ket{7c}$ and $\Ket{8c}$.
The band structure is obtained by solving the eigenvalue equation
\begin{equation}
\label{sec:Theory;eq::2}
	{\Ham}_{14\times14}({\bf k}) \Ket{\lambda,{\bf k}}= \epsilon_{\lambda}({\bf k}) \Ket{\lambda,{\bf k}}
\end{equation}
which is accomplished by a matrix diagonalization. 
The coupling between the lowest conduction band $\Ket{6c}$ and the higher conduction bands $\Ket{7c}$ and $\Ket{8c}$,
\begin{equation}
\label{sec:Theory;eq::3}
\begin{split}
\Ham_{8c 6c} &= -\sqrt{3} P^\prime \left(U_x k_x +cp \right)    \\
\Ham_{7c 6c} &= \frac{1}{\sqrt{3}} P^\prime \left(\sigma_x k_x +cp\right) 
\end{split}\text{,}
\end{equation}
is responsible for the shift current and consequently the respective bands have to be included in the simulations.\cite{Podzimski_SPIE_15}
At $T=0K$ the band gap is $E_0=1.519\,\mathrm{eV}$ while the distance between the valence band $\Ket{8v}$ and the higher conduction band is $E^\prime_0 = 4.488\,\mathrm{eV}$.\cite{Aspnes_PRB_73,Cardona_PRB_88}
Thus when numerically solving the SBE significantly small time steps have to be used to resolve the rapid oscillations arising from these energetic differences of the
involved bands.

The time evolution of the photoexcited system is described by the SBE\cite{Duc_PRB_10,Sternemann_PRB_13,Duc_PRB_2016},
i.e., the Heisenberg equations of motion for $x^{\lambda \lambda^\prime}_{\bf k} = \langle a^{\dagger}_{\lambda{\bf k}} a^{}_{\lambda' {\bf k}} \rangle$
representing the coherences ($\lambda \neq \lambda^\prime$) and the occupations ($\lambda = \lambda^\prime$) of the system in k-space, which read 
\begin{equation}
\label{sec:Theory;eq::4}
\begin{split}
\frac{d}{dt}x^{\lambda \lambda^\prime}_{\bf k} =& \frac{\iu}{\hbar} \left( \epsilon^{\lambda}_{\bf k} -\epsilon^{\lambda^\prime}_{\bf k} \right)x^{\lambda \lambda^\prime}_{\bf k} \\
{}& +\frac{i}{\hbar}  \sum_\mu \left( {\Omega}^{\mu \lambda}_{\bf k}  x^{\mu \lambda^\prime}_{\bf k}  - {\Omega}^{\lambda^\prime \mu}_{\bf k}  x^{\lambda \mu}_{\bf k} \right)
-\frac{1}{T_{1/2}}x^{\lambda \lambda^\prime}_{\bf k}
\text{,}
\end{split}
\end{equation}
with $\epsilon^{\lambda}_{\bf k}$ and $\epsilon^{\lambda^\prime}_{\bf k}$ being the energies of band $\Ket{\lambda}$ and $\Ket{\lambda^\prime}$ at point ${\bf k}$ in reciprocal space.
The generalized Rabi-frequency
\begin{equation}
\label{sec:Theory;eq::5}
{\Omega}^{\lambda \lambda^\prime}_{\bf k} = \frac{e}{m_0}{\bf A}(t)  \cdot {\bf \Pi}^{\lambda \lambda^\prime}_{\bf k} + \sum_{\mu \mu^\prime {\bf q}} V^{\lambda \mu \lambda^\prime \mu^\prime}_{{\bf k},{\bf q}} x^{\mu \mu^\prime}_{{\bf k} +{\bf q}}(t) 
\end{equation}
contains the light-matter interaction, here written in the velocity gauge ${\bf \Pi}\cdot {\bf A}(t)$, and the Coulomb interaction in time-dependent Hartree-Fock approximation.\cite{Duc_PRB_10}
In the velocity gauge the light-matter interaction is described by the electrodynamic vector potential ${\bf A}(t)$ and the momentum matrix elements\cite{Chang_PRB_89,Enders_PRB_95}  
\begin{equation}
\label{sec:Theory;eq::6}
		{\bf \Pi}_{\bf k} ^{\lambda \lambda^\prime} = \frac{m_0}{\hbar} \left< \nabla_{\bf k} H({\bf k})  \right>_{\lambda \lambda^\prime}\text{.}
\end{equation}
The Coulomb matrix element is given by\cite{Duc_PRB_10}
\begin{equation}
\label{sec:Theory;eq::7}
V^{\lambda_1,\lambda_2,\lambda_3,\lambda_4}_{{\bf k},{\bf q}}  = \frac{1}{V} \frac{4\pi V_0}{{\bf q}^2} \BraKet{\lambda_1,{\bf k+q}}{\lambda_4,{\bf k}}\BraKet{\lambda_2,{\bf k}}{\lambda_3,{\bf k+q}}\text{,}
\end{equation}
with $V_0 = \frac{e^2}{4\pi\epsilon\epsilon_0}$, $\epsilon = 12.9$ the dielectric constant of GaAs\cite{Landolt_2001}, and
$V=\frac{1}{(2\pi)^3}$ is the volume of the reciprocal unit cell.

So, the Coulomb matrix elements depend on four band indices ($\lambda_1,\lambda_2,\lambda_3,\lambda_4$) and two three-dimensional k-space vectors (${\bf k},{\bf q}$).
Evaluating the Coulomb matrix for all 14 bands results in $14^4=38416$ possible combinations of band indices.
This number can be drastically reduced by considering only the resonantly excited bands near the band gap, i.e., $\Ket{8v}$ and $\Ket{6c}$, which leaves $6^4=1296$ band index combinations
but still is sufficient to properly describe the excitonic absorption.
Additionally, the Coulomb matrix couples all k-space vectors $\bf k$ and $\bf q$ which leads to a numerical effort when calculating the matrix elements and solving the SBE
which grows quadratically as function of the total number of $\bf k$-points $N_k$.
Using a standard Cartesian grid with $N_k = N^3$, with $N$ being the number of k-points in one direction, results in a numerical effort that grows with $N^6$.
To achieve converged results an unreasonably high memory and computer time would be required.
A favorable alternative to a Cartesian grid is the so-called geodesic grid which is a spherical grid that is often used in climate simulations.\cite{Heikes_MWR_95a,Mahdavi_IJDE_15}

The {\bf k}$\cdot${\bf p} band structure Hamiltonian can written as a sum which distinguishes the involved symmetries\cite{Lipari_PRL_70}
\begin{equation}
\label{sec:Theory;eq::8}
\Ham = \Ham_{sphere} + \Ham_{cube} + \Ham_{tetra}\text{.}
\end{equation}
For small $\bf k$-vectors the spherical ${\bf k}^2$-terms have the largest contribution to the Hamiltonian while for large $\bf k$-vectors higher order $\bf k$-terms representing cubic and tetrahedral symmetry
become more relevant.
Thus at the $\Gamma$-point the energy contributions to the total Hamiltonian can be ordered as\cite{Luttinger_PRW_56}
\begin{equation}
	\label{sec:Theory;eq::9}
	 \Ham_{sphere} > \Ham_{cube} > \Ham_{tetra} \text{.}
\end{equation}
which is the reason why the band structure at the $\Gamma$-point can be reasonable well be approximated by a parabolic dispersion and the analytic solutions of Wannier excitons,
which have a spherical $\Ket{S}$-symmetry known from the hydrogen atom, are in good agreement with experimental results.\cite{Wannier_PR_1937,Luttinger_PRW_56}
The application of a spherical grid takes advantage of the predominant spherical symmetry at the $\Gamma$-point reducing the amount of wasted simulation space and
consequently saving computational effort.
In a spherical grid based on a spherical coordinate system, ${\bf r}(r,\theta,\varphi)$, the point density on a sphere surface is inhomogeneous and increases towards the pols.
This numerically undesirable property of spherical coordinates is circumvented in a geodesic grid in which the sphere's surface is constructed almost completely from equilateral
hexagons and thus has a very homogeneous point density. 
The total number of points in a geodesic grid is 
$N_k = N_A \, N_R$
with $N_A$ being the number of points per sphere and $N_R$ the number of spheres.
Consequently, using the geodesic grid the numerical evaluation of the SBE with Coulomb interaction
can be reduced from a $N^6$ scaling to a more moderate $N_R^2 \, N_A^2$ scaling
which is a huge improvement regarding the required numerical resources.
Further details can be found in \cite{Reinold_diss}.


\section{Numerical Results}
\label{sec::Results}

Here, we present and discuss results on the excitonic absorption and the exciton wave function in Sec.~\ref{sec::Exciton}
and on the shift currents in  Sec.~\ref{sec::Shift}.
In our calculations we use a geodesic grid with $N_R = 60$ spheres and $N_A = 12$ points per sphere and
we thus assign more importance to the radial resolution than to the angular resolution.
This can be justified by the fact that spherical symmetry is the dominant symmetry the of {\bf k}$\cdot${\bf p}-Hamiltonian at the $\Gamma$-point
and that this resolution results in a converged exciton resonance.
For numerical stability we introduce a weak screening and evaluate the Coulomb matrix element as
\begin{equation}
\label{Sec::Spectrum;Eqn::1}
V^{\lambda_1,\lambda_2,\lambda_3,\lambda_4}_{{\bf k},{\bf q}}  = \frac{1}{V} \frac{4\pi V_0}{{\bf q}^2+c_{S}^2} \BraKet{\lambda_1,{\bf k+q}}{\lambda_4,{\bf k}}\BraKet{\lambda_2,{\bf k}}{\lambda_3,{\bf k+q}}\text{.}
\end{equation}
As screening constant we use $c_{S}=4.5\cdot10^{-3}$~nm$^{-1}$ which is much smaller than the inverse of the exciton Bohr radius
$1/a_B \approx 1/12$~nm$=0.083$~nm$^{-1}$ and thus only weakly influences the spectra.
In our simulations we use a dephasing time of $T_2 = 800$~fs which is chosen sufficiently long in order to spectrally separate the exciton peak properly from the band edge.

\begin{figure}[htbp!]
\centering
\includegraphics[width=0.5\textwidth]{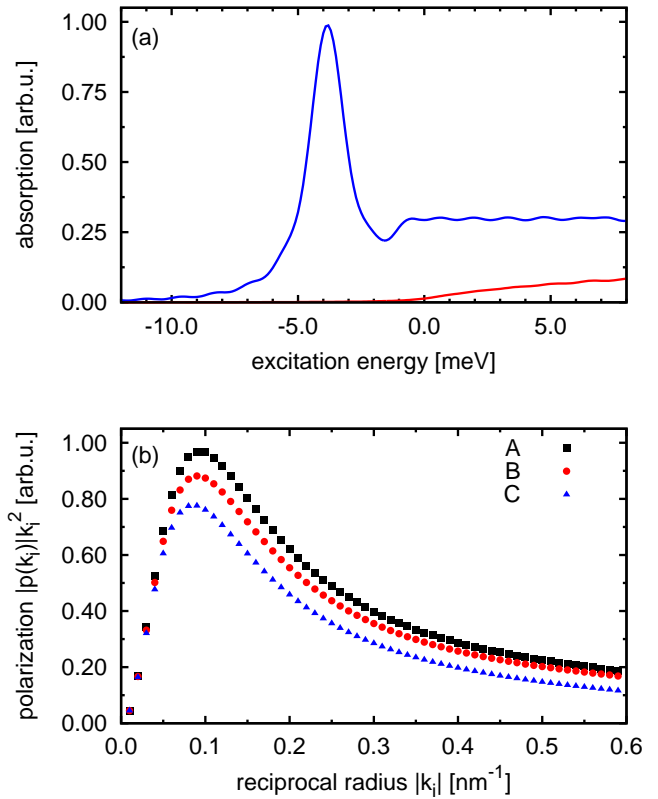}
\caption{(color online) (a) Linear absorption spectra for bulk GaAs calculated with (blue) and without (red) Coulomb interaction.
The excitation energy is defined relative to the low-temperature band gap of GaAs of $E_{gap}=1.519$~eV.
The weak oscillations are caused by the finite integration time of $3$~ps. 
(b) The absolute value of the optically-excited microscopic polarization $k_i^2|p(k_i)|$
as function of the radius $|k_i|$.
The black (A), red (B), and blue (C) symbols represent different directions in k-space  present in the 
geodesic grid as explained in the main text.}
\label{Sec::Exciton;Fig::1}
\end{figure}

\subsection{Linear absorption and exciton wave function}
\label{sec::Exciton}

To calculate linear absorption spectra, we numerically solve the SBE for excitation with a weak linearly polarized ultrashort pulse
and obtain the time-dependent total optical polarization ${\bf P}(t) = \sum_{c v , \bf k}  \frac{i \omega_{\bf k} ^{c v}}{m_0} {\bf \Pi}_{\bf k}^{c v} x^{c v}_{\bf k}  (t)$. 
Here, $\omega_{\bf k} ^{c v}$ is the transition frequency between the valence band $v$ and the conduction band $c$ at wave vector ${\bf k}$
and the term $\frac{i \omega_{\bf k} ^{c v}}{m_0} {\bf \Pi}_{\bf k}^{c v}$ corresponds to the transition dipole matrix element.
The linear absorption is then taken to be proportional to the imaginary part of the Fourier transformed polarization in the direction of the linearly excitation ($x$), i.e.,
$\alpha(\omega)\propto Im[P_x(\omega)]$.
Figure~\ref{Sec::Exciton;Fig::1}(a) shows the calculated absorption spectrum of bulk GaAs with and excitonic effects.
Without Coulomb interaction the absorption starts at the band gap $E_{gap}$ and follows a $\sqrt{E-E_{gap}}$-dependence
which originates from the density of states in three dimensions for a parabolic dispersion.
When excitonic effects are included the absorption spectrum contains a well defined exciton peak at $E_{X} \approx 4.0$~meV,
thus the obtained exciton binding energy is in good agreement with literature values.
Furthermore, Fig.~\ref{Sec::Exciton;Fig::1}(a) shows the Coulomb enhanced continuum absorption which is basically constant for energies slightly above the band gap.
The increased absorption direction below the band gap originates from excited exciton states which are not resolved individually and
merge with the continuum absorption.

Unlike a parabolic band structure, the {\bf k}$\cdot${\bf p} band structure contains anisotropies, e.g.,
small splittings caused by the spin-orbit interaction that appear in the (111) and (100) directions.
To investigate the influence of this anisotropy on the exciton wave function,
we solve the SBE with a vector potential corresponding to a weak and slowly rising electric field, which is given by
\begin{equation}
\label{Sec::Exciton;Eqn::1}
E(t) = 
\left \{
\begin{array}{rl} 
exp \left\{ -\left(\frac{t}{400\,\mathrm{fs}}\right)^2\right\}\sin(\omega_L t), & t < 0 \\ 
\sin(\omega_L t) , & t \ge 0
\end{array}
\right . \text{,}
\end{equation}
where the excitation frequency $\omega_L$ chosen to be resonant with the exciton.
We integrate the SBE up to $t=3$~ps and due to the narrow spectral width of the excitation only the exciton is significantly excited.
Thus for the chosen exciting field, the microscopic polarization in reciprocal space is proportional to the exciton wave function, i.e.,
\begin{equation}
\Psi_{exc}({\bf k }) \propto  | p({\bf k}) |  \text{.}
\end{equation}
The microscopic polarization $p({\bf k})$ is determined as the $x$-component of 
$\frac{i \omega_{\bf k} ^{c v}}{m_0} {\bf \Pi}_{\bf k}^{c v} x^{c v}_{\bf k}$, which is the component in the direction
of the linearly polarized excitation pulse.
Since each point on a sphere of the geodesic grid contributes proportional the square of the radius when integrating
over k-space, we multiply $|p({\bf k})|$ by $k^2$ in Fig.~\ref{Sec::Exciton;Fig::1}(b). 
We see, that the exciton wave function has a peak at approximately $|k| \approx 0.08\,\mathrm{nm^{-1}}$ which corresponds to the inverse Bohr radius $\frac{1}{a_B}$, with $a_B \approx 12$~nm
being the effective Bohr radius of the exciton in GaAs.\cite{chow_book_99}
For an isotropic exciton the wave function does not depend on the direction in k-space and only a single line should be in Fig.~\ref{Sec::Exciton;Fig::1}(b).
In our {\bf k}$\cdot${\bf p} approach, however, the anisotropy of the band structure leads to a dependence of the exciton wave function
on the direction in k-space, as is evidenced by the three different lines visible in Fig.~\ref{Sec::Exciton;Fig::1}(b).
(A) corresponds to the $(0,{+1},{+\varphi})$ in reciprocal space,
where $\varphi = \frac{1+\sqrt{5}}{2} \approx 1.618$ is the so called golden ratio which is used in the construction of the geodesic grid.
Very similar results are also obtained in $(0,{-1},{+\varphi})$, $(0,{+1},{-\varphi})$, and $(0,{-1},{-\varphi})$, so we can write
$(0{,\pm1}{,\pm \varphi})$ as a shorthand notation for these directions.
(B) and (C) correspond to the 4 directions $({\pm1},{\pm \varphi},{0})$ and $({\pm \varphi},{0},{\pm1})$, respectively. 
The difference between the wave function amplitudes in the different directions is significant and amounts to approximately 20\%
and thus clearly demonstrates the anisotropy of the exciton wave function.

\begin{figure}[htbp!]
\centering
\includegraphics[width=0.5\textwidth]{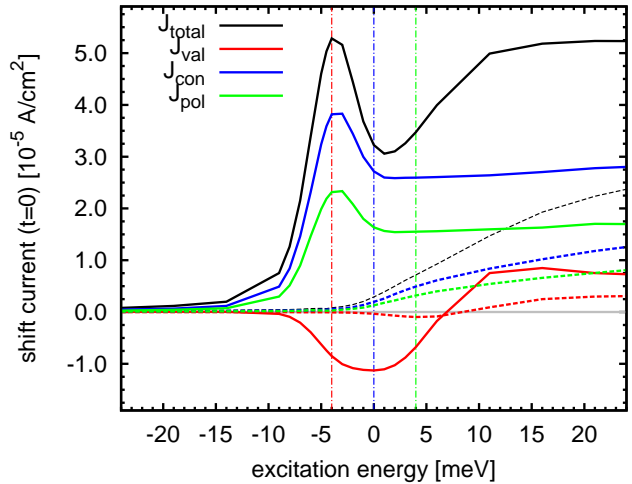}
\caption{(color online) 
Excitation frequency dependent shift currents evaluated at the center of Gaussian excitation pulse.
The full and the dashed lines display simulation results with and without Coulomb interaction, respectively.
The black lines represent the full currents, while the red, blue, and green lines show the sub currents generated by the valence band, the conduction band, and the interband polarization, respectively.
The intensity of the incident pulse is $0.1$~W/cm$^2$ and its duration 500~fs (FWHM of the pulse intensity).
}
\label{Sec::Current;Fig::1}
\end{figure}

\subsection{Excitonic and near-band-gap shift currents}
\label{sec::Shift}

The photocurrents induced by optical interband excitation
are determined from solutions of the SBE by\cite{Duc_PRB_2016} 
\begin{equation}
\label{sec:Theory;eq::11}
{\bf J} (t) = \sum_{\lambda, \lambda^\prime \neq \lambda , {\bf k}} {\bf \Pi}^{\lambda  \lambda^\prime}_{\bf k} x^{\lambda \lambda^\prime}_{\bf k}\text{.}
\end{equation}
The shift current is a $\chi^{(2)}(0;\omega,-\omega)$ response and as such represents the $\omega\approx 0$ components of Eq.~\eqref{sec:Theory;eq::11}.
Therefore ${\bf J}(t)$ is Fourier transformed and to the resulting ${\bf J}(\omega)$
a frequency filter is applied around $\omega \approx 0$.
Considering a filter function $F(\omega)$ which cuts off frequency components with $\hbar \omega \ge 100$~meV,
we obtain ${\bf J}_{shift}(\omega)=F(\omega){\bf J}(\omega)$.
By Fourier transforming back to the time domain we finally obtain ${\bf J}_{shift}(t)$.
As ${\bf J} (t)$ is given by a double summation over the 14 bands, i.e., $\sum_{\lambda \lambda^\prime}$,
one can distinguish different contributions arising from transitions between different
valence and conduction bands.
Thus one can write ${\bf J}_{shift} (t)$, as a sum
\begin{equation}
\label{sec:Theory;eq::11x}
{\bf J}_{shift} (t) =
{\bf J}_{val} (t) +
{\bf J}_{con} (t) +
{\bf J}_{pol} (t)
\text{,}
\end{equation}
where ${\bf J}_{val}$ (${\bf J}_{con}$) originates from intersubband transitions between different valence (conduction) bands
and 
${\bf J}_{pol}$ originates from interband transitions between a valence and a conduction band.

In Fig.~\ref{Sec::Current;Fig::1}(a) we show shift currents with and without Coulomb interaction
which are generated by excitation
with a pulse with a Gaussian envelope of 500~fs duration (FWHM of the pulse intensity)
at the center of the pulse as function of the excitation frequency.
The pulse is linearly polarized in (110)-direction which leads to a shift current $J^z(t)$ in (001)-direction.
Without Coulomb interaction the shift current vanishes for excitation frequencies below the band gap and
follows the square root like behavior of the three-dimensional density of states for above band gap excitation.
With Coulomb interaction the shift current is strongly enhanced.
At $E-E_{gap} \approx -4.0\,\mathrm{meV}$ a peak is visible in the total shift current which corresponds to excitation of the exciton resonance. 
For excitations energies above the band gap the total current initially grows with excitation energy and then approaches a basically constant value.

Analyzing the different sub currents, introduced above, offers an explanation for this behavior.
$J_{con}$ and $J_{pol}$, the shift currents created by the conduction band and the interband polarization, have both a peak at the exciton and remain mostly constant for above band gap.
$J_{val}$, the valence band shift current, displays a very different behavior.
At the exciton energy $J_{val}$ displays a negative current and remains negative up until $E-E_{gap} \approx 7\,\mathrm{meV}$, where the current changes its direction and then changes to a positive current.
The reason for this behavior is two-fold.
For excitation at the exciton and at the band gap the holes are correlated with the electrons and in average flow in the same directions.
Due to the opposite sign of their charges, the resulting signs of corresponding currents are opposite.
For excitation energies higher in the band the holes are less correlated with the electrons and therefore, as without Coulomb interaction,
both currents have a positive sign, which corresponds to electrons and holes moving in opposite directions.
In addition to the shift current contribution, $J_{val}$ also includes the intersubband coherence between the heavy hole and the light hole band.
Since both bands are degenerate at the $\Gamma$-point, their energy difference starts with $\Delta \epsilon_{hh-lh} = 0$ and thus cannot be removed from the valence band shift
current via Fourier filtering.
This coherence is also present without Coulomb interaction and leads to a very small negative current when exciting at the band gap in $J_{val}$.

\begin{figure}[htbp!]
\centering
\includegraphics[width=0.5\textwidth]{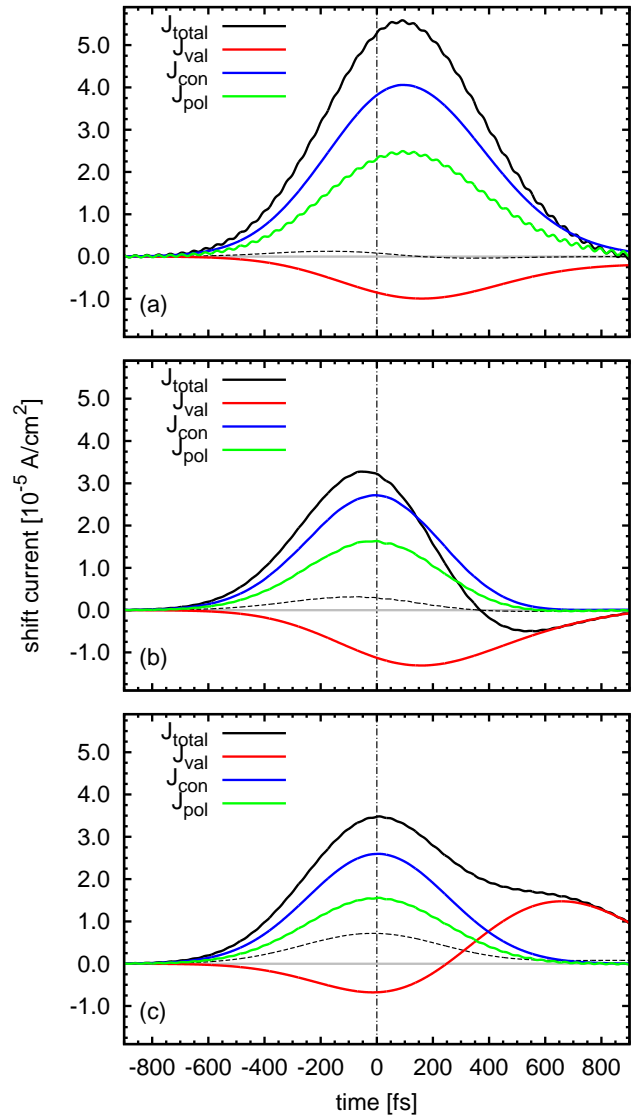}
\caption{(color online) 
Temporal dynamics of the total shift current and its sub currents for excitation energies of (a) $-4$~meV, (b) $0$~meV, and (c) $4$~meV, respectively,
with vertical dashed-dotted lines marking the center of the excitation pulse ($t=0$) at which the current values shown in Fig.~\ref{Sec::Current;Fig::1} are taken.
The dashed black line is the total shift current without Coulomb interaction for the respective excitation energies.
The weak oscillations that appear in some transients are a numerical artifact arising from the finite integration time
due to the Fourier filtering procedure.
}
\label{Sec::Current;Fig::2}
\end{figure}

The calculated time evolution of shift currents after excitation with a pulse with a Gaussian envelope of 500~fs duration
that has its maximum at $t=0$ is shown in Fig.~\ref{Sec::Current;Fig::2}.
For excitation at the band gap, see Fig.~\ref{Sec::Current;Fig::1}(b), $J_{con}$ and $J_{pol}$ have a Gaussian shape, centered at $t \approx 0$.
This is to be expected since the shift current involves is an off-resonant excitation process and in the limit of a purely off-resonant excitation the signal should strictly follow
the envelope of the optical excitation $E_{env}^2(t)$\cite{Nastos_PRB_06}.
This behavior is also in agreement with previous results obtained with the SBE\cite{Podzimski_SPIE_15,Podzimski_SPIE_16,Duc_PRB_2016}
where excitonic effects have been neglected.
$J_{val}$, however, shows a weak temporal delay to positive times which originates from optically induced coherences between the
heavy- and light-hole bands.
When exciting $4\,\mathrm{meV}$ above the band gap, see Fig.~\ref{Sec::Current;Fig::1}(c), $J_{con}$ and $J_{pol}$ remain basically unchanged.
$J_{val}$, however, clearly displays a slow oscillatory behavior which is caused by the now dominant influence of
the coherences between the heavy- and light-hole bands whose average energetic separation increases with increasing distance from the
$\Gamma$-point and thus with excitation energy.

When exciting at the exciton resonance, see Fig.~\ref{Sec::Current;Fig::1}(a), the total current $J_{total}$ and the sub currents have basically a Gaussian shape.
In this case, however, a small delay of about $100\,\mathrm{fs}$ to positive times is visible.
This temporal shift does not significantly change when we use different dephasing times in our simulations.
It can be interpreted by the following effect: When the excitation is tuned at or above the band gap, already the linear optical polarization
corresponds to a superposition of a continuum of transitions with different frequencies which interfere destructively
and thus lead to a decay of the total linear polarization on the time scale of the exciting pulse. 
This situation is different when we tune the excitation to the exciton resonance which is a single discrete optical transition
whose linear polarization increases with the integral over the envelope of the exciting pulse, at least as long as dephasing is neglected.
The above explained difference in the dynamics of the linear optical polarization depending on the spectral position of the
excitation could be responsible for the delayed maximum of the excitonic shift current seen in Fig.~\ref{Sec::Current;Fig::1}(a).

Nastos and Sipe introduced the concept of the shift distance\cite{Nastos_PRB_06}
\begin{equation}
\label{Sec::Current;eq::1}
	{\bf d}_{shift} = \frac{{\bf J}_{shift}}{e \frac{dn}{dt}} \, ,
\end{equation}
where $n$ is the photoexcited density,
which describes the average distance that the electrons shift in space when optically excited from the valence to the conduction band.
Eq.~\eqref{Sec::Current;eq::1} was derived in a frequency-domain approach considering continuous wave excitation.
Our time-domain simulations model transient situation with pulsed excitation and dephasing and relaxation of the coherences and occupations
which makes Eq.~\eqref{Sec::Current;eq::1} rather unsuitable.
Based on simple electrodynamic consideration we define a time-dependent shift distance by
\begin{equation}
\label{Sec::Current;eq::2}
	{\bf d}_{shift}(t) = \frac{\int^t_{-\infty} dt^\prime {\bf J}_{shift}(t^\prime)}{e n(t)} \, .
\end{equation}
Due to the pulsed excitation and the dephasing and relaxation processes considered in our approach Eq.~\eqref{Sec::Current;eq::2}
can only approximate the shift distance and in particular for long times ${\bf d}_{shift}(t)$ unphysically increases
since $n(t)$ decreases.
In our numerical evaluations we find a plateau of the time-dependent shift distance during early times of the excitations\cite{Reinold_diss}
and therefore determine ${\bf d}_{shift}$ in the (001) direction using Eq.~\eqref{Sec::Current;eq::2}
by averaging over the rising part of the incident optical pulse using the time interval $-800$~fs to $-200$~fs.
For exciting at the exciton we obtain a shift distance of $d_{shift}(-4\,\mathrm{meV}) \approx 89\,\mathrm{pm}$.
When shifting the photon energy to above the band gap,
the shift distance increases, e.g.,
 $d_{shift}(10\,\mathrm{meV}) \approx 98\,\mathrm{pm}$ and
 $d_{shift}(31\,\mathrm{meV}) \approx 103\,\mathrm{pm}$.
This agrees with the interpretation that the attraction between holes and electrons causes a reduction of the shift distance for excitations near the exciton.
Without Coulomb interaction we obtain a shift distance $d_{shift} \approx 200\,\mathrm{pm}$  which is quite close to
the average shift distance of $d_{shift} \approx 250\,\mathrm{pm}$ 
calculated in \cite{Nastos_PRB_06} for GaAs considering
continuous wave excitation high above the band gap.
Since the 14 band {\bf k}$\cdot${\bf p} band structure is not well suited to describe
{\bf k} vectors far away from the $\Gamma$-point, we, however,
cannot analyze excitations very high above the band gap
and thus not directly compare with the results of \cite{Nastos_PRB_06}.


\section{Conclusions}
\label{sec::Conclusion}

Using the semiconductor Bloch equations in a basis obtained from a 14 band {\bf k}$\cdot${\bf p} model of the band structure
we calculate linear absorption spectra including excitonic effects and shift current transients of bulk GaAs
for near band gap excitation.
The electron-hole Coulomb attraction results in an excitonic resonance with a binding energy which is in good agreement
with other approaches and experiments.
We show that the anisotropy of the band structure leads to a significant anisotropy of the exciton wave function.
The calculated frequency-dependence of the shift current has a peak at the exciton resonance.
However, the enhancement at the exciton is weaker than the excitonic enhancement in the linear absorption
and correspondingly the shift distance when exciting at the exciton is weaker compared to excitations above the band gap.
The reduced excitonic shift distance as well as the sign change of the valence band current for below band excitation 
originates from the electron-hole attraction.
In addition, for above band gap excitations coherences between the heavy- and light-hole bands lead to oscillatory 
signature in the current transients.
Our findings demonstrate that our approach is able to provide important information on the ultrafast electron dynamics
on the atomic scale.
It is also applicable to study recently predicted Berry-curvature-induced currents arising from bound excitons \cite{ExBerry}
which we plan to investigate for pulsed excitation in future work.

\section{Acknowledgments}
We are grateful to John Sipe for stimulating and fruitful discussions.
This work is supported by the Deutsche Forschungsgemeinschaft (DFG) through the projects ME 1916/2 and SFB/TRR 142 (Project A02).
H.~T. Duc acknowledges the financial support of the Vietnam National Foundation for Science and Technology Development (NAFOSTED) under Grant No.
103.01-2017.42.
We thank the PC$^2$ (Paderborn Center for Parallel Computing) for providing computing time.

\bibliography{shift_coul_bib} 

\end{document}